\documentclass[fleqn,usenatbib]{mnras}

\usepackage{newtxtext,newtxmath}

\usepackage[T1]{fontenc}

\DeclareRobustCommand{\VAN}[3]{#2}
\let\VANthebibliography\thebibliography
\def\thebibliography{\DeclareRobustCommand{\VAN}[3]{##3}\VANthebibliography}

\usepackage{graphicx, float, hyperref, tabularx, physics, ulem, multirow}	
\usepackage{amsmath}	
\usepackage{amssymb}	

\usepackage{threeparttable} 

\newcommand{\maxi}{{MAXI~J1535$-$571}}
\newcommand{\rxte}{{\it{RXTE}}}
\newcommand{\hxmt}{{\textit{Insight}-HXMT}}
\newcommand{\nicer}{{\it{NICER}}}

\title[Corona of \maxi]{A \nicer\ look at the jet-like corona of \maxi\ through type-B quasi-periodic oscillations}

\author[Y.\ Zhang et al.]{
Yuexin Zhang,$^{1}$\thanks{E-mail: yzhang@astro.rug.nl}
Mariano M\'{e}ndez,$^{1}$\thanks{E-mail: mariano@astro.rug.nl}
Federico Garc\'{i}a,$^{2}$
Diego Altamirano,$^{3}$
Tomaso M.\ Belloni,$^{4}$
\newauthor
Kevin Alabarta,$^{5,6}$
Liang Zhang,$^{7}$
Candela Bellavita,$^{8}$
Divya Rawat,$^{9}$
and Ruican Ma$^{1,7,10}$
\\
$^{1}$Kapteyn Astronomical Institute, University of Groningen, P.O. BOX 800, 9700 AV Groningen, The Netherlands\\
$^{2}$Instituto Argentino de Radioastronom\'{i}a (CCT La Plata, CONICET; CICPBA; UNLP), C.C.5, (1894) Villa Elisa, Buenos Aires, Argentina\\
$^{3}$School of Physics and Astronomy, University of Southampton, Southampton, SO17 1BJ, UK\\
$^{4}$INAF-Osservatorio Astronomico di Brera, via E.\ Bianchi 46, I-23807 Merate, Italy\\
$^{5}$Center for Astro, Particle and Planetary Physics, New York University Abu Dhabi, PO Box 129188, Abu Dhabi, United Arab Emirates\\
$^{6}$New York University Abu Dhabi, PO Box 129188, Abu Dhabi, United Arab Emirates\\
$^{7}$Key Laboratory of Particle Astrophysics, Institute of High Energy Physics, Chinese Academy of Sciences, Beijing 100049, People's Republic of China\\
$^{8}$Facultad de Ciencias Astron\'{o}micas y Geof\'{i}sicas, Universidad Nacional de La Plata, Paseo del Bosque, B1900FWA La Plata, Argentina\\
$^{9}$Inter-University Centre for Astronomy and Astrophysics, Ganeshkhind, Pune 411007, India\\
$^{10}$University of Chinese Academy of Sciences, Chinese Academy of Sciences, Beijing 100049, People's Republic of China
}

\date{Accepted XXX. Received YYY; in original form ZZZ}

\begin{document}
\label{firstpage}
\pagerange{\pageref{firstpage}--\pageref{lastpage}}
\maketitle

\begin{abstract}
\maxi\ is a black-hole X-ray binary that in 2017 exhibited a very bright outburst which reached a peak flux of up to 5~Crab in the 2--20~keV band. Given the high flux, several X-ray space observatories obtained unprecedented high signal-to-noise data of key parts of the outburst. In our previous paper we studied the corona of \maxi\ in the hard-intermediate state (HIMS) with \hxmt. In this paper we focus on the study of the corona in the soft-intermediate state (SIMS) through the spectral-timing analysis of 26 \nicer\ detections of the type-B quasi-periodic oscillations (QPOs). From simultaneous fits of the energy, rms and lag spectra of these QPOs with our time-dependent Comptonization model, we find that in the SIMS the corona size is $\sim$ 6500~km and vertically extended. We detect a narrow iron line in the energy spectra, which we interpret to be due to the illumination of the outer part of the accretion disk by this large corona. We follow the evolution of the corona and the radio jet during the HIMS-SIMS transition, and find that the jet flux peaks after the time when the corona extends to its maximum vertical size. The jet flux starts to decay after the corona contracts vertically towards the black hole. This behavior points to a connection between the X-ray corona and the radio jet similar to that seen in other sources.
\end{abstract}




\section{Introduction}\label{sec:intro}

Black-hole low-mass X-ray binaries (BH LMXBs) consist of a stellar-mass black hole with a low-mass companion star overflowing its Roche lobe, forming an accretion disk~\citep[see][for reviews]{2011BASI...39..409B,2022arXiv220614410K}. During an outburst, a black-hole X-ray transient traces an anticlockwise `q' path in the hardness-intensity diagram~\citep[HID;][]{2005Ap&SS.300..107H,2005A&A...440..207B}, during which the proportion of the thermal and Comptonized components changes leading to several well-defined spectral states~\citep[][]{2005A&A...440..207B}. The accretion disk emits thermal emission in the soft X-ray band with the disk spectrum peaking at around 0.2--2.0~keV; a fraction of the soft disk photons are Compton up-scattered in the corona, forming a power-law like component in the energy spectrum in the hard X-ray band up to 100~keV~\citep[for reviews, see][]{2007A&ARv..15....1D,2010LNP...794...17G}. When the outburst starts, the X-ray luminosity of the source increases by several orders of magnitude compared to the quiescent state and the source enters the low-hard state (LHS) with a dominant hard Comptonized component and a relatively weak thermal disk component in the X-ray spectrum~\citep[e.g.][]{2018MNRAS.481.5560S,2018ApJ...855...61W}. As the mass accretion rate increases, the source becomes brighter and quickly goes into the hard-intermediate state (HIMS), the soft-intermediate state (SIMS), the high-soft state (HSS), and sometimes an anomalous state at the highest luminosity~\citep[e.g.][]{1997ApJ...479..926M,2005A&A...440..207B,2010LNP...794...53B,2012MNRAS.427..595M}. From the LHS to the HSS, the truncated disk moves closer to the innermost circular stable orbit (ISCO) around the black hole (\citealt{1997ApJ...489..865E}; but see~\citealt{2008MNRAS.387.1489R}; \citealt{2018ApJ...855...61W} and \citealt{2018ApJ...860L..28M} for a different interpretation) and the thermal component becomes dominant, while the spectrum of the hard component becomes steeper and weaker~\citep[e.g.][]{2018MNRAS.481.5560S,2022MNRAS.514.1422D}. With the gradual decay of the mass accretion rate, the source evolves back to the LHS and finally returns to the quiescent state. In some so-called failed-transition outbursts the source stays in the LHS and HIMS and never enters the SIMS~\citep{2021MNRAS.507.5507A}.

The Comptonized photons in the corona can impinge back onto the accretion disk, resulting in the reprocessing and Compton back-scattering of the hard photons, which leads to a relativistic reflection component~\citep{1989MNRAS.238..729F,2021SSRv..217...65B}. The reflection spectrum includes characteristic X-ray emission lines among which the most prominent is the iron $K_{\alpha}$ line around 6.4--7.0~keV and the Compton hump around 20~keV~\citep{2014ApJ...782...76G}. Recent studies have also modeled the relativistic reflection considering the incident photons coming from a hot blackbody or a returning reflection spectrum due to the strong gravitational bending~\citep[e.g.][]{2020ApJ...892...47C,2022ApJ...926...13G,2022MNRAS.514.3965D}. A relativistic reflection component can appear through the spectral evolution from the LHS to the HSS~\citep[e.g.][]{2018ApJ...855...61W,2020ApJ...892...47C,2022MNRAS.514.1422D}.

The X-ray light curves of black-hole binaries (BHBs) show variability on time scales from milliseconds to years~\citep[see][for a recent review]{2019NewAR..8501524I}. In the Fourier domain, the power density spectrum (PDS) of the X-ray light curve shows different kinds of variability, e.g.\ broadband noise and narrow peaks called quasi-periodic oscillations~\citep[QPOs;][]{1989ASIC..262...27V,2000MNRAS.318..361N,2002ApJ...572..392B}. The QPOs observed in BHBs are classified as low-frequency QPOs (LFQPOs), with central frequencies ranging from mHz to $\sim$ 30~Hz, and high-frequency QPOs (HFQPOs), with central frequencies $\gtrsim$ 60~Hz~\citep{2014SSRv..183...43B}. Depending on the shape of the broadband noise, the fractional root-mean-square amplitude (hereafter rms), the phase lags of the QPOs and the spectral state of the source, LFQPOs are divided into three classes, namely type A, B, and C~\citep{2002ApJ...564..962R,2004A&A...426..587C,2005ApJ...629..403C}. Type-C QPOs generally appear in the hard states (both LHS and HIMS) with central frequency in the $\sim$ 0.01--30~Hz range, high rms (up to 20\%), strong band-limited noise, and usually second and sub harmonics. Type-B QPOs only appear in the SIMS~\citep{2016ASSL..440...61B}, with central frequency below $\sim$ 6~Hz, low rms ($\lesssim 5\%$), weak red noise, and sometimes the second and sub harmonics. Type-A QPOs, which are rarely detected, appear in the SIMS and HSS, with central frequencies in the $\sim$ 6--8~Hz range, very weak rms and no harmonics. In the short-lived SIMS, fast transitions are sometimes observed between the type-B and either other types of QPOs or the disappearance of QPO~\citep[e.g.][]{2004A&A...426..587C,2021MNRAS.505.3823Z}.

The Fourier cross spectrum of two simultaneous light curves in different energy bands (subject and reference band) can be used to compute the phase lags as a function of Fourier frequency. It is common to report lags of the signals identified in the PDS extending over a given frequency range, e.g. the broadband noise and the QPOs~\citep{1999ApJ...510..874N}. Hard (positive) lags~\citep{1988Natur.336..450M} can be produced by propagation of fluctuations of the mass accretion from the outer part towards the inner part of the disk and the corona~\citep[e.g.][]{2006MNRAS.367..801A,2013MNRAS.434.1476I}. Soft (negative) lags can be due to reverberation or thermalization of hard photons when the corona photons impinge back onto the accretion disk~\citep[e.g.][]{2014A&ARv..22...72U,2020MNRAS.492.1399K}.

Apart from the X-ray emission, BHBs radio emission from a jet is sometimes prominent, and can be classified into two types depending on the radio spectral index and the morphology of the jet: a small-scale, optically thick, compact jet and an extended, optically thin, transient jet~\citep[for a review, see][]{2006csxs.book..381F}. The relation between the spectral states and the radio emission indicates the existence of an accretion-ejection coupling. In the LHS and the HIMS, a compact jet is observed, while during the state transition from the HIMS to the SIMS, the compact jet can be quenched for a few days~\citep{2004MNRAS.355.1105F}. In the SIMS, there is no longer a compact jet but a bright transient jet appears with observable discrete relativistic ejecta, while in the HSS, the jet disappears~\citep[e.g.][]{1994Natur.371...46M,2004ApJ...617.1272C,2019MNRAS.489.3927I}. Time variability in the radio band sometimes appears, but much less often than in the X-ray band~\citep[e.g.][]{2019MNRAS.484.2987T,2021MNRAS.504.3862T}. Type-B QPOs in the X-ray band are usually thought to be connected to the relativistic transient jet but the exact mechanisms that explain the connection are still unknown~\citep[e.g.][]{2009MNRAS.396.1370F,2020ApJ...891L..29H,2021MNRAS.501.3173G}.

There are still many open questions regarding the the disk-corona-jet evolution, for instance, the disk truncation during the evolution of the black hole transients~\citep{1997ApJ...489..865E}, the nature of the corona~\citep[the inner hot flow, disk sandwich, or the base of the jet;][]{1979ApJ...229..318G,1991ApJ...380L..51H,2005ApJ...635.1203M}, and the geometry of the corona and its typical size~\citep{1996MNRAS.282L..53M,1998MNRAS.299..479L}. The geometry of the corona and its connection with the disk and the jet are still to be understood. A universal radio--X-ray correlation in the LHS provides evidence that the corona can be related to the radio jet~\citep[e.g.][]{2003MNRAS.344...60G,2004MNRAS.355.1105F}. Using data of GRS 1915+105, \citet{2022NatAs...6..577M} proposed that (part of) the spread in the radio--X-ray correlation~\citep{2012MNRAS.423..590G} could be due to changes of the corona temperature. Studies of the spectral energy distribution (SED) from the radio to the X-ray band shows that the corona emission can originate from a shock acceleration region of tens of gravitational radii ($R_{g}$) as the jet base~\citep{2001A&A...372L..25M,2019MNRAS.485.3696C,2022MNRAS.509.2517C}. The spectral analysis of the reflection component in MAXI~J1820+070 suggests that the corona outflows with a higher relativistic velocity as it is closer to the black hole~\citep{2021NatCo..12.1025Y}. Through X-ray variability studies, the size of the corona shows a continuous 
evolution during the outburst that could be related to the change of the radio jet emission, suggesting a disk-corona-jet connection~\citep{2019Natur.565..198K,2021ApJ...910L...3W,2022NatAs...6..577M,2022MNRAS.512.2686Z}.

Many corona models of time variability have been proposed to explain the corona geometry. Using the broadband noise, \texttt{reltrans} models the reverberation lags and measures the corona height assuming a lamppost geometry of the corona~\citep{2019MNRAS.488..324I}. Assuming the corona is a wide, low-velocity, wind-like structure, a corona outflow model is proposed to explain the observed correlations of, for instance, the power-law photon index and time lags and the photon index and radio flux~\citep[][]{2008A&A...489..481K,2018A&A...614L...5K}. The propagation of mass accretion rate fluctuations model assumes that the corona lies in a truncated disk. This model follows the variability (both QPOs and broadband noise) and also fits the time-averaged energy spectrum~\citep[e.g.][]{2013MNRAS.434.1476I,2021ApJ...909L...9Z,2022MNRAS.511..536K}. The JED-SAD model assumes that the hard part of the spectrum comes from a jet-emitting disk (JED) and the soft part from a standard accretion disk (SAD)~\citep{1997A&A...319..340F,2008MNRAS.385L..88P,2018A&A...615A..57M}. The JED-SAD model not only explains the `q' path in the HID in terms of transitions between accretion modes, but also matches the observed variability like the LFQPOs and the hard-soft lags~\citep{2019A&A...626A.115M,2020A&A...640A..18M}. The dynamical origin of the LFQPOs can be explained by the Lense-Thirring (L-T) precession of the corona which also restricts the corona region within a truncated disk or a jet base~\citep{1998ApJ...492L..59S,2009MNRAS.397L.101I,2018ApJ...858...82Y,2021NatAs...5...94M}, or instabilities in the disk accretion flow~\citep{1999A&A...349.1003T}.

Recently \citet{2020MNRAS.492.1399K} and \citet{2022MNRAS.515.2099B} developed a time-dependent Comptonization model called \texttt{vKompthdk} that explains the radiative properties (rms and phase lags) of QPOs and measures the corona geometry around the black hole. The \texttt{vKompthdk} model assumes that the temperatures of the disk and the corona and the rate at which the corona is heated up oscillate coherently at the frequency of the QPO. The unspecified external heating source is required to keep the temperature of the corona that, otherwise, would cool down very quickly and become undetectable. In this model, a hot spherical corona partially covers the soft disk and scatters the seed photons from the disk into the Comptonized hard photons. Note that we assume that the temperature of the corona is constant with radius~\citep{1980A&A....86..121S}. While inverse-Comptonization scattering in the corona may lead to a radial dependence of the corona temperature, this effect is likely negligible, especially at a large radii~\citep{2012A&A...544A..87M}. Part of the out-going hard photons are observed while the others feedback onto the soft disk, are reprocessed, and finally reach thermal equilibrium with the disk. The steady state energy spectrum produced by \texttt{vKompthdk} is the same as that of \texttt{nthcomp}~\citep{1996MNRAS.283..193Z,1999MNRAS.309..561Z}. The inverse-Compton scattering of the soft photons in the corona results in hard lags while the reprocessing of the hard photons in the disk leads to soft lags, both of which can be reproduced by \texttt{vKompthdk} to predict the size of the corona. We note that the model defines the flux of the feedback hard photons divided by the flux of the total Comptonized photons as an intrinsic feedback fraction, $\eta_{\text{int}}$. Apart from the $\eta_{\text{int}}$, in the model an explicit feedback fraction parameter, $\eta$, is the flux of the feedback hard photons divided by the flux of the observed soft disk, thus $\eta$ is in the range of 0--1. The intrinsic feedback fraction indicates the efficiency of hard photons that feedback onto the disk. Combining the measurements of the corona size and the feedback fraction, we can understand to what extent the corona covers the disk and whether the shape of the corona is sphere-like or jet-like~\citep{2021MNRAS.503.5522K,2021MNRAS.501.3173G,2022NatAs...6..577M,2022MNRAS.512.2686Z}.

\maxi\ is an X-ray transient discovered by \textit{MAXI}/GSC and \textit{Swift}/BAT independently when it went into outburst on 2017 September 2~\citep{2017ATel10700....1K,2017ATel10699....1N}. The source reached a peak flux of up to 5~Crab in the 2--20~keV band~\citep{2017ATel10729....1N}. It has been proposed that \maxi\ has a rapidly spinning (> 0.84) black hole and subtends a high inclination angle (> 60$^{\circ}$) through the modeling of the relativistic reflection component~\citep{2018ApJ...860L..28M,2018ApJ...852L..34X,2022MNRAS.514.1422D}. \citet{2022MNRAS.514.1422D} also reported that the corona temperature increases from 18~keV to > 300~keV as the source evolves from the LHS to the SIMS. In the soft state, the luminosity of \maxi\ is near the Eddington luminosity and the structure of the standard disk is likely to become slim~\citep{2018MNRAS.480.4443T}. Timing studies of \maxi\ show that through the outburst, different types of LFQPOs appear and the LFQPOs evolve as the spectral state changes~\citep{2018ApJ...866..122H,2018ApJ...865L..15S}. \citet{2019MNRAS.488..720B} showed a correlation between the frequency of type-C QPOs and the hard photon index, indicating a connection between the timing features and the spectral parameters. From the radio observations of \maxi, the jet is first a compact jet in the HIMS, then quenches during the transition from the HIMS to the SIMS, and finally a transient jet appears in the SIMS~\citep{2019ApJ...883..198R,2020MNRAS.498.5772R,2019MNRAS.488L.129C}.

In this paper, we continue our previous study of the corona geometry of \maxi\ and the connection between the X-ray corona and the radio jet through the type-C QPOs in the HIMS using \hxmt\ observations~\citep{2022MNRAS.512.2686Z}. We further explore the corona properties in the SIMS through the type-B QPOs, which are weak and much less frequently detected, using \nicer\ observations. We fit jointly the rms and phase-lag spectra of the type-B QPO and the time-averaged energy spectra of the source using the latest version of the time-dependent Comptonization model \texttt{vKompthdk}~\citep{2022MNRAS.515.2099B}. This paper is organized as follows: In section~\ref{sec:analysis} we describe the data reduction of the \nicer\ observations of \maxi, and explain how we calculate the rms and phase lags of the type-B QPO in different energy bands. We also explain the parameter settings of the model used to fit jointly the rms and phase-lag spectra of the QPO and the time-averaged energy spectra of the source. In section~\ref{sec:results} we show the X-ray temporal evolution of \maxi, the rms and the phase-lag spectra of the identified type-B QPO, and the spectral parameters measured from the joint spectral fitting. Finally, in section~\ref{sec:discussion} we discuss our results and compare them with previous studies of type-B QPOs and the corona geometry. In that section, we combine the properties of the corona measured in this work with our previous results and propose a more complete picture of the corona-jet connection during the whole evolution of \maxi\ from the HIMS to the SIMS.

\section{Observations and data analysis}\label{sec:analysis}

The \textit{Neutron Star Interior Composition Explorer}~\citep[\nicer;][]{2016SPIE.9905E..1HG} observed \maxi\ from 2017 September 7 to 2019 May 11 for a total of 219 observations. We use the standard \nicer\ reduction routine \texttt{nicerl2} with CALDB version xti20210707 to process the data. We remove the data of detectors \# 14 and \# 34 which are affected by episodes of increased electronic noise. We require the pointing direction of the instrument to be less than 0.015$^{\circ}$ offset, at least 40$^{\circ}$ above the bright Earth limb, and at least 30$^{\circ}$ above the Earth limb. For a bright source like \maxi, we apply an undershoot count rate range 0--200 per module (\texttt{underonly range}). We set the column types of the prefilter to be NICERV3 and 3C50 as the recommended background columns. We require each GTI to be longer than 16~s and split the observations into separate orbits.

\subsection{Light curve and hardness-intensity diagram}

We focus on the 36 observations in the period MJD 58008--58036, i.e.\ the outburst before the four reflares~\citep{2020MNRAS.496.1001C}. We use XSELECT to extract light curves at 1-s resolution for each orbit in the 1--3~keV and the 3--10~keV bands. We exclude the data below 1~keV since we find that the interstellar absorption towards the source is very high (see subsection~\ref{subsec:init fit}), and there is no significantly detected emission from the source below that energy. In order to obtain the HID, for each orbit we use the average count rate in the 1--10~keV band and calculate the hardness ratio using the ratio of the average count rate in the 3--10~keV band and the 1--3~keV band.

\subsection{Energy spectra}\label{subsec:espect}

We extract the energy spectra of the source and the background in each orbit using the \texttt{nibackgen3C50} tool~\citep{2022AJ....163..130R}, and use \texttt{nicerarf} and \texttt{nicerrmf} to generate the ancillary files and the response files, respectively. We group the spectrum such that we have at least 25 counts per bin and we oversample the intrinsic resolution of the instrument by a factor of 3.

We fit the energy spectra using XSPEC v12.12.1~\citep{1996ASPC..101...17A}. For the time-averaged energy spectrum of \maxi, following~\citet{2018ApJ...860L..28M} we use the 2.3--10 keV energy band in order to avoid the calibration in the Si band (1.7--2.1 keV) and the Au band (2.2--2.3 keV), but we notice the energy range between 2.1 keV and 2.2 keV. A systematic error of 0.5\% is added in quadrature to the model when performing the spectral fitting.

We apply the model \texttt{TBfeo*(diskbb+vKompthdk+gaussian)}, hereafter Model~1, to fit the energy spectra. The component \texttt{TBfeo} models the Galactic absorption towards the source with variable oxygen and iron abundances. We set the cross section and the solar abundance of the ISM using the tables of~\citet{1996ApJ...465..487V} and~\citet{2000ApJ...542..914W}, respectively. The component \texttt{diskbb}~\citep{1984PASJ...36..741M} represents the emission of a multi-temperature optically thick and geometrically thin disk with parameters being the disk temperature, $kT_{\text{in}}$, and a normalization. The time-averaged or steady-state version of the time-dependent Comptonization model \texttt{vKompthdk}~\citep{2020MNRAS.492.1399K,2022MNRAS.515.2099B} is the same as \texttt{nthcomp}~\citep{1996MNRAS.283..193Z,1999MNRAS.309..561Z}. The parameters of this component are the seed photon temperature, $kT_{\text{s}}$, the corona temperature, $kT_{\text{e}}$, the photon index, $\Gamma$, and a normalization. The optical depth, $\tau$, is a function of $kT_{\text{e}}$ and $\Gamma$ in \texttt{vKompthdk}:
\begin{equation}
    \tau = \sqrt{\frac{9}{4} + \frac{3}{\left(kT_{\text{e}}/m_{\text{e}}c^2\right)\left(\left(\Gamma + 1/2\right)^2 - 9/4\right)}} - \frac{3}{2}, \label{tau}
\end{equation}
where $m_{\text{e}}$ and $c$ are the mass of electron and the speed of light, respectively, and the Compton \textit{y-parameter}~\citep{1969SvA....13..175Z,1976ApJ...204..187S}, $y = \text{max}(\tau, \tau^2)\, 4kT_{\text{e}} / (m_{\text{e}}c^2)$, drives the shape of the spectrum.
The seed photon temperature, $kT_{\text{s}}$, in \texttt{vKompthdk} is linked to the inner disk temperature, $kT_{\text{in}}$, in \texttt{diskbb}. In fact \texttt{vKompthdk} contains four extra parameters that only affect the time-dependent spectrum produced by this model, and describe the corona size, $L$, the feedback fraction, $\eta$, the amplitude of the variability of the rate at which the corona is heated by an (unspecified) external source, $\delta\dot{H}_{\text{ext}}$, and an additive parameter, reflag, that gives the phase lag in the 2--3~keV band\footnote{Since in the data the reference band used to compute the lags is arbitrary, the lags are defined up to an additive constant.}. Note that none of these four parameters changes the steady-state spectrum produced by \texttt{vKompthdk}. These four parameters, plus $kT_{\text{s}}$, $kT_{\text{e}}$, and $\Gamma$, describe the radiative properties of the QPOs, i.e. the rms and the phase lags. We add a Gaussian component to fit the broad iron $K_{\alpha}$ emission line in the spectrum. We do not use more complicated reflection models like \texttt{relxill}, since in the energy band of \nicer\ we only detect the iron line feature and we do not have data above 10~keV where the Compton hump appears.

\subsection{Power spectrum}

We generate the PDS for each orbit in the 1--10~keV energy band and in five narrow bands, 1--2.5, 2.5--3.9, 3.9--5.5, 5.5--7.3, 7.3--10.0~keV bands. The length of each PDS segment is 8.192~s and the Nyquist frequency is 125~Hz. In each orbit we average all the PDS segments, subtract the Poisson noise using the average power from 100~Hz to 125~Hz in the PDS and normalize the PDS to fractional rms amplitude~\citep{1990A&A...230..103B}:
\begin{equation}
    \text{rms} = \frac{\sqrt{P(S+B)}}{S}, \label{eq:rms}
\end{equation}
where $P$ is the power in Leahy units~\citep{1983ApJ...266..160L}, and $S$ and $B$ are, respectively, the source and background count rates. We apply a logarithmic rebin in frequency to the PDS such that the bin size increases by a factor $\exp(1/100)$ compared to the previous one.

We fit the averaged PDS of each orbit in the 1--10~keV band with Lorentzian functions~\citep{2000MNRAS.318..361N,2002ApJ...572..392B} in the frequency range of 0.1--30~Hz. The parameters of a Lorentzian function are the central frequency, the full width at half maximum (FWHM), and the normalization. We fit the PDS with four Lorentzians representing the QPO fundamental, the second harmonic, and two broadband noise components. An extra Lorentzian is needed if there is a QPO sub harmonic. All parameters in all the Lorentzians are free, but we freeze the central frequency of one Lorentzian function which fits the broadband noise at 0. If there is no QPO, we reduce the number of Lorentzian functions to one or two to only fit the broadband noise. Using the models that we apply to fit the PDS of each orbit in the 1--10~keV band as baselines, we further fit the PDS of each orbit in the five narrow energy bands (see above). We fix all central frequencies and the FWHMs of all the Lorentzians to the values we obtained for the PDS in the 1--10~keV band and only let the normalizations free to fit the PDS in those five narrow energy bands. We also check that the central frequency and width of those Lorentzians do not change significantly with energy. Finally we take the square root of the normalizations of the Lorentzians that represent the variability components to calculate the rms.

\subsection{Phase lag spectrum}~\label{subsec:plag}

We generate and average the Fast Fourier Transformation (FFT) in each orbit using the 1--10~keV energy band as reference band to compute the phase lags. The length of each FFT segment is 8.192~s and the Nyquist frequency is 125~Hz. We also compute the FFT of the data in the energy bands 1--2.5~keV, 2.5--3.9~keV, 3.9--5.5~keV, 5.5--7.3~keV, 7.3--10.0~keV that we use as the subject bands to compute the cross spectrum for each subject band with respect to the reference band.

Considering that the cross spectrum as a function of Fourier frequencies is $G(f) = \text{Re}\, G(f) + i\, \text{Im}\, G(f)$, where the phase lag is $\phi(f) = \tan^{-1}\left(\text{Im}\,G(f)/\text{Re}\,G(f)\right)$, we can calculate the phase lags of the variability components that we measure in the PDS~\citep{1997ApJ...474L..43V,1999ApJ...510..874N,2019MNRAS.489.3927I}. As in~\citet{2022MNRAS.513.2804P} and~\citet{2022MNRAS.514.2839A}, we perform a simultaneous fitting of both the real and the imaginary parts of $G(f)$ using the same Lorentzian components that we used to fit the PDS. We add a constant in the model to fit the correlated part of the signal that arises because the subject bands are always within the 1-10~keV reference band~\citep{2019MNRAS.489.3927I}. This constant is not linked between the real and the imaginary parts of the cross spectra. We fix the central frequency and the FWHM of all the Lorentzians at the same values that we get from fitting the PDS. For the real part, we fit the normalizations of the Lorentzians, while for the imaginary part, we define an extra free parameter, $\phi$, for each Lorentzian so that the normalization of a Lorentzian in the imaginary part is equal to $\tan{\phi}$ times the normalization of the corresponding Lorentzian in the real part. We note that this assumes that the phase lag of each component is constant with frequency~\citep{2022MNRAS.513.2804P}. Using this method, for the energy bands we defined, we compute all the phase lags of all the variability components fitted by the Lorentzian functions.

\section{Results}\label{sec:results}

\subsection{Light curve and hardness-intensity diagram}\label{subsec:hid}

\begin{figure*}
    \includegraphics[width=0.99\textwidth]{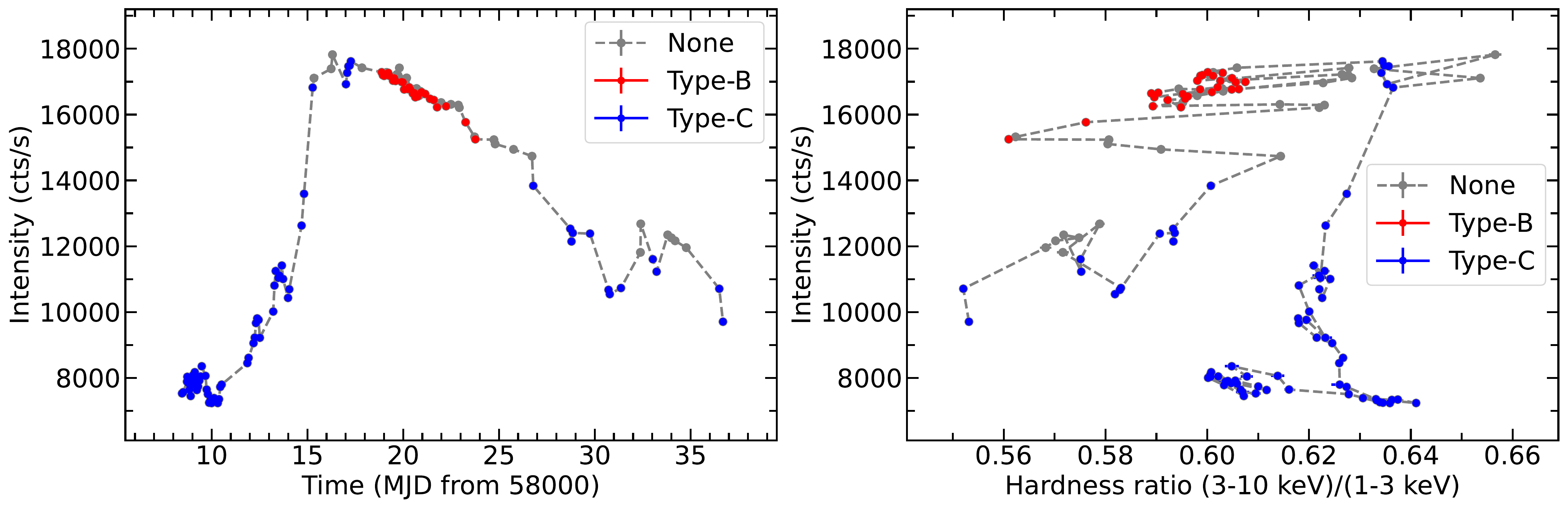}
    \caption{Light curve (left) and hardness-intensity diagram (right) of \maxi\ during the main outburst from MJD~58008 to 58037. Each point corresponds to a \nicer\ orbit. The grey, red, and blue points indicate observations with, respectively, no QPO, type-B QPOs, and type-C QPOs. The error bars indicating 68\% confidence level are smaller than the size of the points. The grey dashed lines connect the data points in time sequence.}
    \label{fig:hid}
\end{figure*}

Fig.~\ref{fig:hid} shows the light curve (left) and the HID (right) of \maxi\ in the time period MJD~58008--58036. Each data point corresponds to one \nicer\ orbit. During the outburst, the source moves in an anticlockwise direction in the HID. From MJD 58008 to 58011, \maxi\ stays in the LHS with a count rate $\sim$ 8000~cts/s and a hardness ratio increasing from $\sim$ 0.6 to $\sim$ 0.64. From MJD 58011 to 58017 the count rate increases quickly from $\sim$ 7000~cts/s to $\sim$ 18000~cts/s while the hardness ratio increases slowly from $\sim$ 0.63 to $\sim$ 0.64 with some excursions up to a maximum of 0.66. At this point the source moves to the left in the HID and transits from the LHS to the HIMS~\citep[e.g.][]{2005A&A...440..207B}. From MJD 58017 to 58024 the count rate gradually decreases to $\sim$ 15000~cts/s, while the hardness ratio decreases from $\sim$ 0.62 to $\sim$ 0.56, indicating a state transition from the HIMS to a softer state. During the final part of the outburst, from MJD~58024 to 58036, the count rate continues decreasing gradually to 10000~cts/s, while the hardness ratio decreases from 0.6 to 0.56.

We mark with blue in Fig.~\ref{fig:hid} the observations where we detect the type-C QPOs. The data of the type-C QPOs are from~\citet{2023arXiv230104418R} who performed a systematic study of the type-C QPOs in the \nicer\ observations of \maxi. The observations mark the hard and the hard-intermediate states, and do not include the observations of the soft-intermediate and the soft states where weak QPOs may appear~\citep[e.g.][for a review]{2014SSRv..183...43B}. In the top part of the HID (Fig.~\ref{fig:hid}) we plot in grey observations with no significant QPO in the PDS. The remainder of the orbits, 26 in total, show a QPO with a rather constant frequency of $\sim$ 5--6~Hz. The information of the 26 orbits is listed in Tab.~\ref{tab:obs}. We plot these observations in Fig.~\ref{fig:hid} with red. Since these QPOs in the SIMS are weak and their central frequencies are within a small range, to improve the signal-to-noise ratio we average the PDS of the 26 orbits. The left panel of Fig.~\ref{fig:pds} shows the averaged PDS. Based on the fact that both the QPO fundamental and the second harmonic appear and given the weak rms of the zero-centered broadband noise (0.7\%; see Tab.~\ref{tab:concatenate_pds_par}), we tentatively identify these QPOs as type-B QPOs.

Since the red points in the HID (the right panel of Fig.~\ref{fig:hid}) extend to a very soft region with hardness ratio less than 0.59, some of the QPOs we detected may be type-A~\citep[e.g.][]{2004A&A...426..587C}. To check whether all our detections are type-B QPOs we divide the observations into two groups with hardness ratio either greater or smaller than 0.59. We find that in both groups the averaged PDS are consistent with the PDS shown in Fig.~\ref{fig:pds}. We therefore conclude that all the QPOs that appear in the observations marked with red in Fig.~\ref{fig:hid} are type-B QPOs.

\subsection{Initial fitting to the time-averaged energy spectra}\label{subsec:init fit}

We first fit all the energy spectra of the 26 orbits with the type-B QPOs separately using the model \texttt{TBfeo*(diskbb+vKompthdk+gaussian)}, as introduced in subsection~\ref{subsec:espect}. The best-fitting values of $\Gamma$, $kT_{\text{in}}$ and the iron line profiles in the separate orbits are consistent with being the same within errors and only the normalizations of the Comptonized and the disk components change. Given that $kT_{\text{e}}$ cannot be constrained using the \nicer\ spectra,  we fix it to 250~keV.

Since the only parameters that change for the different orbits are the normalizations of the disk and the corona components, to improve the signal-to-noise ratio (SNR) we fit the energy spectra of the 26 orbits together, using the same model \texttt{TBfeo*(diskbb+vKompthdk+gaussian)} linking all the parameters in each of the 26 data groups, except for the normalizations of \texttt{diskbb} and \texttt{vKompthdk} components. Note that this fitting can also make it more convenient for our later joint fitting in subsection~\ref{subsec:joint}. This spectral fitting shows that, at this stage, the electron temperature of the corona $kT_{\text{e}}$ still cannot be constrained, therefore we fix it again to 250~keV. We find that the best-fitting hydrogen column density is $4.2\times 10^{22}$ $\text{cm}^{-2}$, consistent with previous measurements~\citep{2018ApJ...860L..28M}. The disk temperature is 1.14~keV, while the corona photon index is 2.84, indicating that the source is in a relatively soft state~\citep[][]{2022MNRAS.514.1422D}. The \texttt{gaussian} used to fit the iron line has a central energy of 6.8~keV with a width of 0.98~keV. For the 26 orbits, the 0.5--10~keV averaged flux of the disk and the Comptonized components is, respectively, $1.6\times 10^{-7}$~erg cm$^{-2}$ s$^{-1}$ and $0.4\times 10^{-7}$~erg cm$^{-2}$ s$^{-1}$.

\subsection{Fractional rms and phase-lag spectra}

\begin{figure*}
    \includegraphics[width=0.99\textwidth]{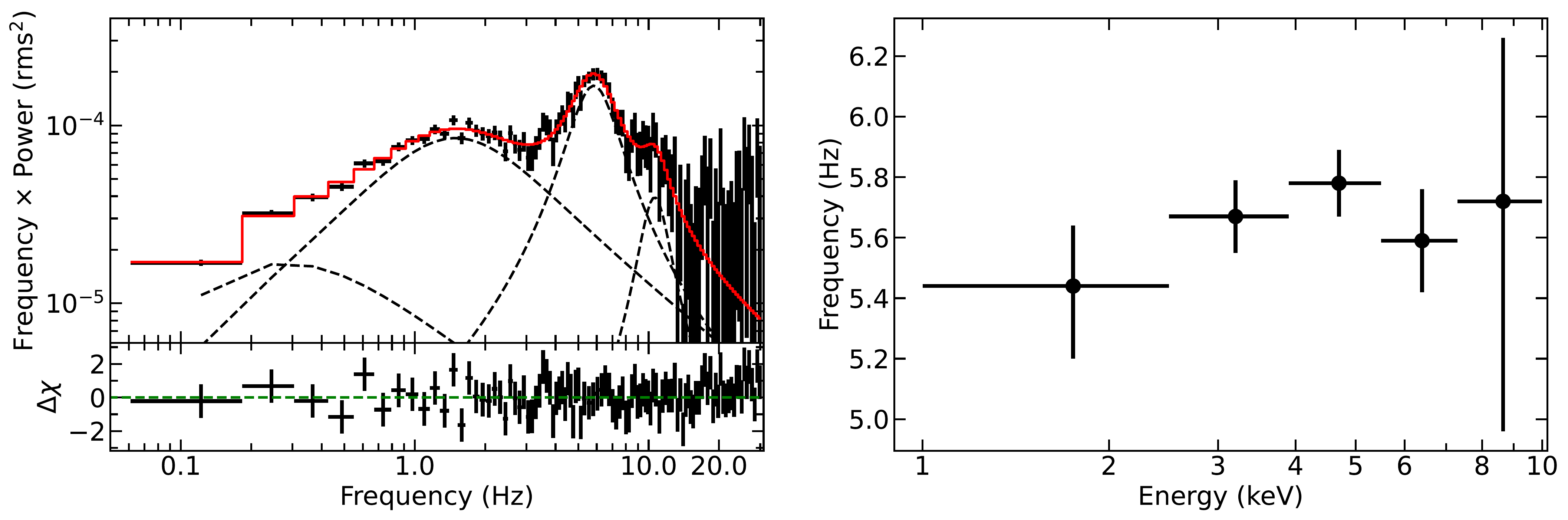}
    \caption{Left panel: Averaged PDS of \maxi\ in the 1--10~keV band for the \nicer\ orbits listed in Tab.~\ref{tab:obs}, with the best-fitting model and the residuals. The black points are the data and the red line is the best-fitting model. The dashed lines from the left to the right show two broadband noise, the type-B QPO fundamental and the second harmonic components. The residuals are the data minus the model divided by the error. Right panel: The QPO central frequency vs. energy for the type-B QPO. The error bars indicate the 68\% confidence level.} 
    \label{fig:pds}
\end{figure*}

\begin{figure*}
    \includegraphics[width=0.99\textwidth]{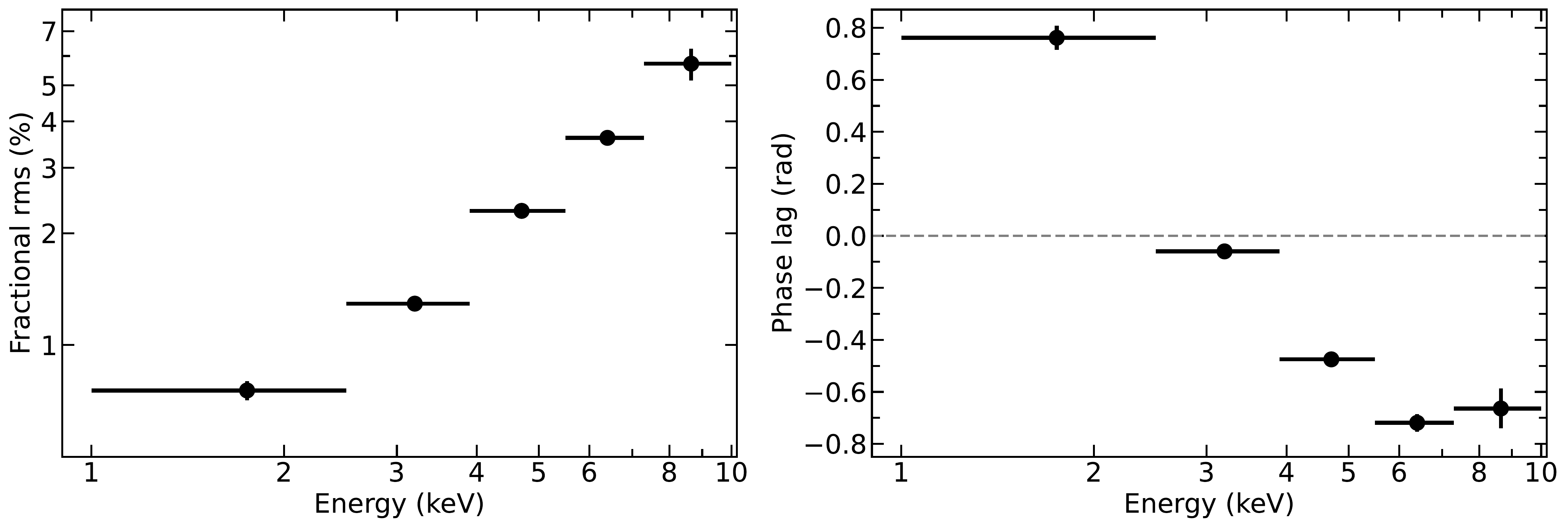}
    \caption{Fractional rms amplitude spectrum (left panel) and phase-lag spectrum (right panel) of the type-B QPO of \maxi. The dashed grey line in the right panel marks the zero phase lag. In both panels, the vertical error bars indicates the 68\% confidence level, while the horizontal error bars indicate the width of the energy channels.} 
    \label{fig:rms_and_plag_spect}
\end{figure*}

As shown in the left panel of Fig.~\ref{fig:pds}, we fit the PDS with four Lorentzians representing two broadband noise components, the QPO fundamental, and the second harmonic. Tab.~\ref{tab:concatenate_pds_par} gives the best-fitting parameters of the Lorentzians. The average type-B QPO has a central frequency of $5.64\pm 0.07$~Hz with a FWHM of $2.9\pm 0.3$~Hz. The type-B QPO is weak, with an rms of $1.1\pm 0.05\%$. The second harmonic of the type-B QPO appears at a central frequency of $10.5\pm 0.5$~Hz, consistent with being twice the central frequency of the QPO fundamental. 

We plot the rms spectrum of the QPO in the left panel of Fig.~\ref{fig:rms_and_plag_spect}. The horizontal error bars are the width of the energy channels. As the energy increases from 1~keV to 10~keV, the rms increases monotonically from $\sim 1\%$ to $\sim 6\%$, indicating that the type-B QPO is mainly modulated by the hot corona. 

The right panel of Fig.~\ref{fig:rms_and_plag_spect} shows the phase-lag spectrum of the type-B QPO. The horizontal error bars are the width of the energy channels. As the energy increases from 1~keV to 7.3~keV, the phase lags decrease from 0.8~rad to -0.7~rad, while above 7.3~keV the phase lags increase slightly to around -0.6 rad, resulting in a minimum in the phase-lag spectrum at around 6--7~keV. The energy where the minimum phase lag appears is consistent with the results of~\citet{2018ApJ...865L..15S}. The phase lag with respect to the the 1--10~keV reference band crosses the zero line (grey dashed line in the right panel of Fig.~\ref{fig:rms_and_plag_spect}) from the positive to the negative at an energy of $\sim$ 3~keV.

\begin{table}\renewcommand\arraystretch{1.2}
    \caption{The parameters of the Lorentzians used to fit the averaged PDS of \maxi. The error bar indicates the 1-$\sigma$ confidence level.}
    \centering
    \begin{threeparttable}
        \begin{tabularx}{\columnwidth}{ccccc} 
            \hline
            Component & Frequency (Hz) & FWHM (Hz) & rms (\%) \\
            \hline
            QPO & $5.64\pm 0.07$ & $2.9\pm 0.3$ & $1.11\pm 0.05$ \\
            Harmonic & $10.5\pm 0.5$ & $3\pm 2$ & $0.43\pm 0.08$ \\
            BBN1$^{*}$ & $0$ & $0.5\pm 0.1$ & $0.73\pm 0.08$ \\
            BBN2$^{*}$ & $0.8\pm 0.1$ & $2.5\pm 0.2$ & $1.40\pm 0.07$ \\
            \hline
        \end{tabularx}\label{tab:concatenate_pds_par}
        \begin{tablenotes}
            \item[*] BBN: Broadband noise component.
        \end{tablenotes}
    \end{threeparttable}
\end{table}

\subsection{Joint fitting of the rms and phase-lag spectra of the QPO and the energy spectra of the source}\label{subsec:joint}

Similar to what we did in~\citet{2022MNRAS.512.2686Z}, we fit jointly the rms and phase-lag spectra of the QPO and the energy spectra of the 26 orbits of \maxi\ (the latter are initially fitted in subsection~\ref{subsec:init fit}) using the model \texttt{TBfeo*(diskbb+vKompthdk+gaussian)}.

Note that for the fitting of the rms spectrum, \texttt{vKompthdk} is multiplied by a \texttt{dilution} component which is not explicitly written in the total model. The introduction of the \texttt{dilution} model is justified since the time-dependent Comptonization model computes the rms spectrum of a variable corona but the observed rms is reduced by all the non-variable spectral components. Here, we assume that the \texttt{diskbb} and the \texttt{gaussian} components are not variable, and therefore the \texttt{dilution} component is $\text{Flux}_{\text{Compt}}(E) / \text{Flux}_{\text{Total}}(E)$ such that $\text{rms}_{\text{Obs}} = \text{rms}_{\text{Compt}} * \text{Flux}_{\text{Compt}} / \text{Flux}_{\text{Total}}$. (Note that this \texttt{dilution} component does not introduce any new parameters to the fits.)

\begin{figure*}
    \includegraphics[width=0.99\textwidth]{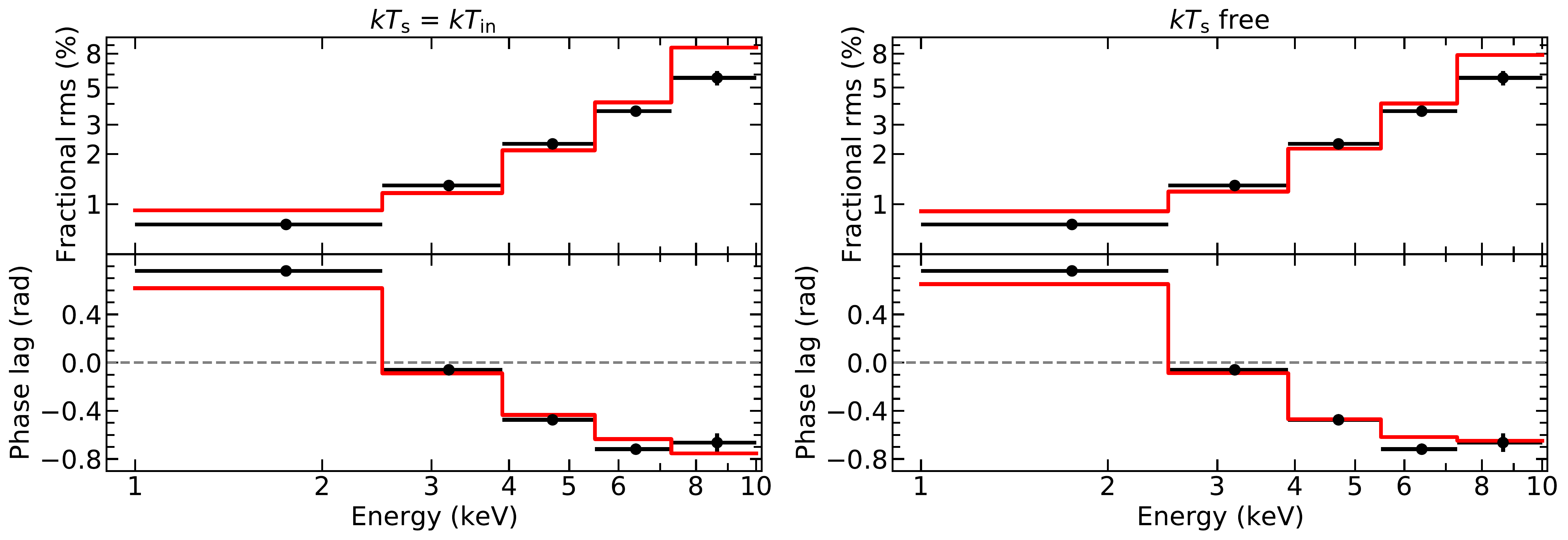}
    \caption{A comparison of the best-fitting model to the rms spectra (upper panels) and phase-lag spectra (lower panels) in the case in which $kT_{\text{s}}=kT_{\text{in}}=1.141$~keV (left panels) and $kT_{\text{s}}$ is free (right panels). In the case in which $kT_{\text{s}}=kT_{\text{in}}$, $\chi^{2}=104.0$ for 10 bins, while in the case in which $kT_{\text{s}}$ is free, $\chi^{2}=62.3$ for 10 bins. The data and the best-fitting model are in black and red, respectively.} 
    \label{fig:comparison_free_kTs}
\end{figure*}

Initially, when we fit simultaneously the energy spectra of the 26 orbits and the rms and the phase-lag spectra of the QPO, $kT_{\text{s}}$ in \texttt{vKompthdk} is linked to $kT_{\text{in}}$ in \texttt{diskbb}. The fit yields a relatively large $\chi^{2} = 104.0$ using 10 bins for the timing spectra and a corona size of $\sim 10^4$~km. The large corona size indicates that the softer seed photons from the outer part of the disk play an important role in producing the phase lags. Therefore, we let $kT_{\text{s}}$ in \texttt{vKompthdk} linked with $kT_{\text{in}}$ in \texttt{diskbb} to fit the energy spectra of the 26 orbits, but allow $kT_{\text{s}}$ to vary freely when fitting the rms and phase-lag spectra. We show a comparison of the best-fitting results of the cases with $kT_{\text{s}}=kT_{\text{in}}$ and $kT_{\text{s}}$ free in Fig.~\ref{fig:comparison_free_kTs}. Letting $kT_{\text{s}}$ free gives a significantly better fit, $\chi^{2} = 62.3$ (Tab.~\ref{tab:spectral_par}), with $kT_{\text{s}} = 0.59\pm 0.07$~keV, lower than $kT_{\text{in}} = 1.139\pm 0.002$~keV; the corona is still relatively large with a size of $6500\pm 500$~km. In the time-averaged energy spectrum $kT_{\text{in}}$ and $\Gamma$ are the same as when $kT_{\text{s}}=kT_{\text{in}}$. The feedback fraction pegs at the upper boundary 1, which gives $\eta_{\text{int}} = 0.33\pm 0.02$. (For an explanation of the difference between the $\eta$ and $\eta_{\text{int}}$, see section~\ref{sec:intro}.) This means that $\sim 33\%$ of the corona photons return to the accretion disk where they are thermalized and re-emitted, resulting in a soft lag, while the other 67\% of the corona photons are the observed Comptonized photons. The temperature of the inner disk, $kT_{\text{in}} = 1.139\pm 0.002$~keV, and the photon index, $\Gamma=2.91\pm 0.06$, indicate that the source is in a relatively soft state. The spectral state matches well with the fact that the type-B QPO appears in the SIMS. Since we now include the rms and phase-lag spectra of the QPO and let $kT_{\text{s}}$ free in the fits, we try to let $kT_{\text{e}}$ free. From the joint fitting, we measure a hot corona with temperature $kT_{\text{e}} = 330\pm 50$~keV, which is consistent with the corona temperature measured by~\citet{2022MNRAS.514.1422D} in the SIMS of \maxi\ using the broadband data of \textit{NuSTAR} and \hxmt.

From the values of $kT_{\text{e}}$ and $\Gamma$ from Model~2, the optical depth of the corona is $\tau=0.16$ and the Compton \textit{y-parameter} is $y=0.41$, indicating that the system is in the unsaturated Comptonization regime. This is consistent with the assumptions of the time-dependent Comptonization model \texttt{vKompthdk} that we used.

We note that, although a fit with a dual corona~\citep[see][]{2021MNRAS.501.3173G} may reduce further the $\chi^{2}$, we do not explore it here because of the limited number of energy bands in the rms and phase-lag spectra of the QPO and the large number of free parameters that would be involved.

\begin{figure*}
    \includegraphics[width=0.99\textwidth]{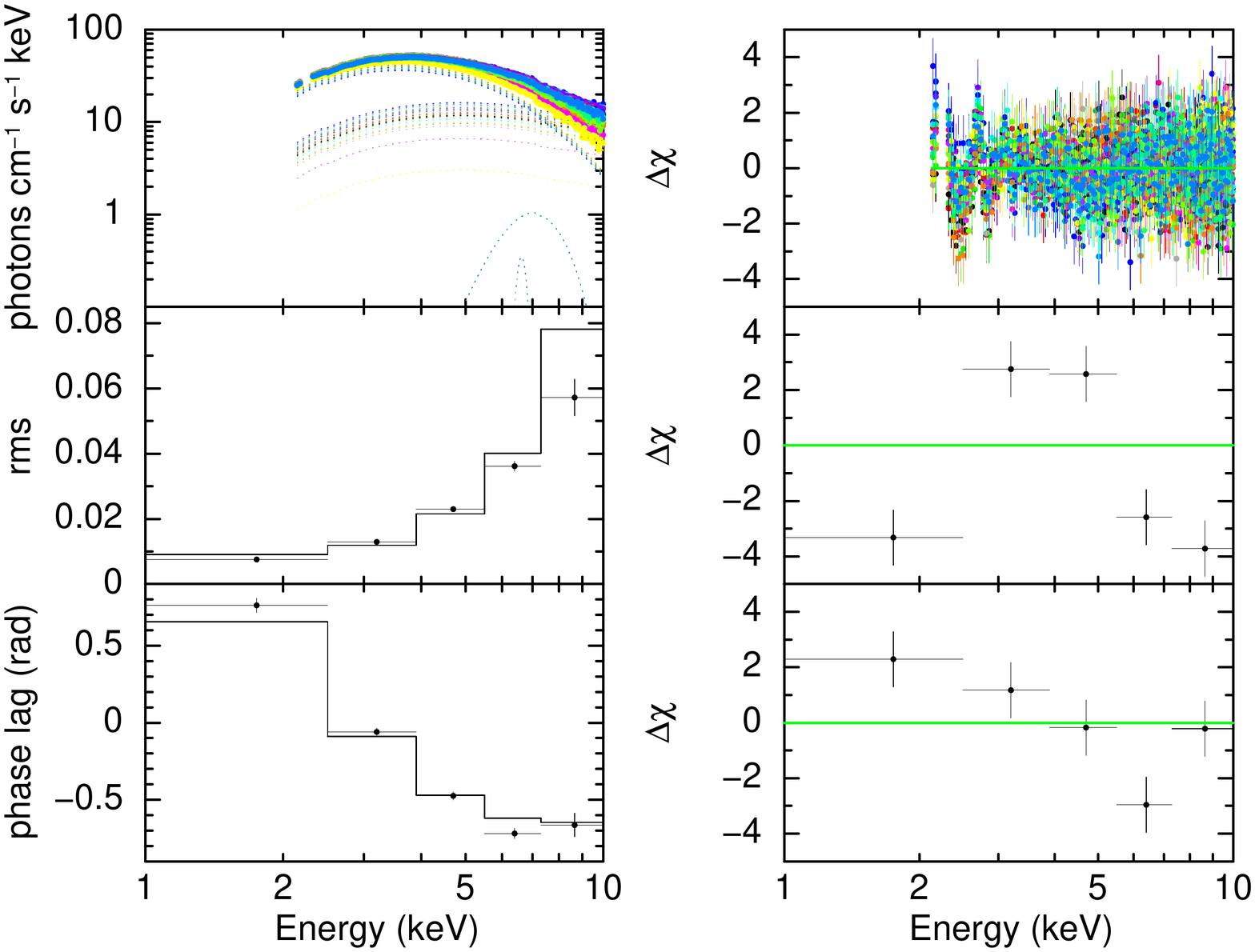}
    \caption{The joint fitting of \maxi\ using \nicer\ data. From top to bottom, the left panels show, respectively, the energy spectra of the 26 orbits, and the rms and phase-lag spectra of the type-B QPO with the best-fitting models. The right panels show the residuals with respect to the best-fitting model.} 
    \label{fig:joint_fitting}
\end{figure*}

Since the corona is large, it should illuminate the outer part of the disk and produce a narrow iron line due to reflection off cold material there~\citep[e.g.][]{2018ApJ...860L..28M}. Therefore, we add an extra Gaussian line to Model 1, which then becomes Model 2: \texttt{TBfeo*(diskbb+vKompthdk+gaussian1+gaussian2)}. We find that a second Gaussian line centered at $6.62\pm 0.02$~keV with a width of $0.12\pm 0.02$~keV fits the data well. The significance of the narrow Gaussian line, measured as the ratio of the line normalization to its error, is 5 $\sigma$ and an F-test yields a probability of $1.8\times 10^{-9}$, which indicates that the narrow Gaussian line improves the fit significantly. The final results of the fitting are plotted in Fig.~\ref{fig:joint_fitting} and the parameters are given in Tab.~\ref{tab:spectral_par}.

\begin{figure*}
    \includegraphics[width=0.99\textwidth]{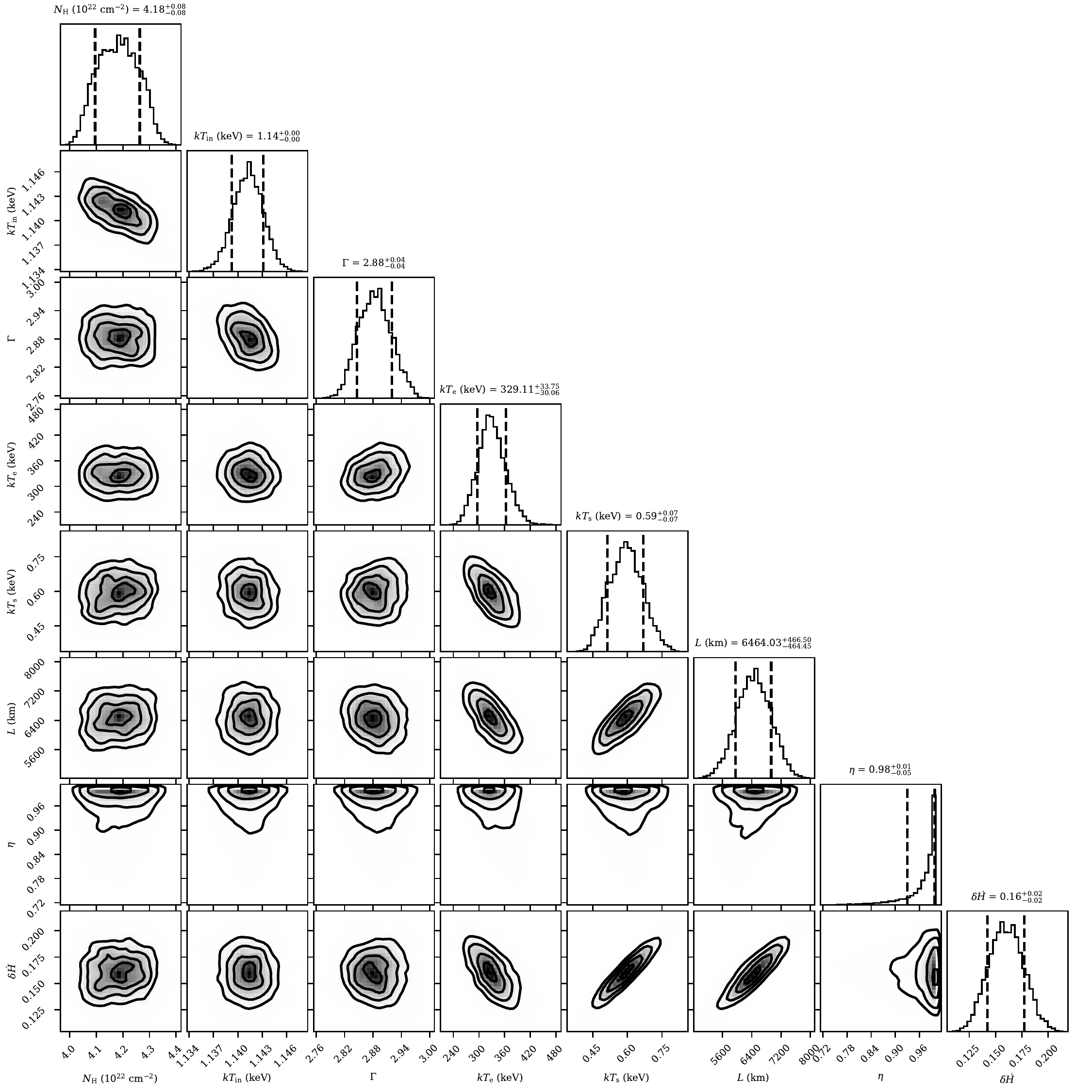}
    \caption{MCMC simulation of the spectral parameters of \maxi. The contours in each panel indicate the 1-$\sigma$, 2-$\sigma$ and 3-$\sigma$ confidence ranges. The parameters are the same as those in the Model~2 in Tab.~\ref{tab:spectral_par}.} 
    \label{fig:mcmc}
\end{figure*}

To test a possible degeneracy of the parameters, we run an MCMC simulation for Model 2, using the Goodman-Weare chain algorithm~\citep{2010CAMCS...5...65G}. After testing the convergence of the chain, we set the chain length to 240000 with 1200 walkers. We discard the first 120000 steps and record the last 120000 steps. The entire covariance matrix is divided by a factor of 10000 to ensure that the walkers initially sample a large range of the parameter space. Fig.~\ref{fig:mcmc} shows the results of the MCMC simulation. The inner disk temperature, $kT_{\text{in}}$, is somewhat covariant with the Galactic absorption, $N_{\text{H}}$, which is expected since the absorption influences mainly the soft part of the spectrum, while the outer disk temperature, $kT_{\text{s}}$, is covariant with the corona temperature, $kT_{\text{e}}$, since these two temperatures play an important role when computing the phase lags~\citep{2022MNRAS.515.2099B}. For the corona geometry, the size of the corona may be modulated by the seed photons from the outer disk and the hot corona temperature. The feedback fraction reaches its upper limit and a 3-$\sigma$ error also constrain the feedback fraction to be larger than 0.9. Since the external heating rate maintains the temperature of the system when describing the timing spectra, $\delta\dot{H}_{\text{ext}}$ is correlated with the outer disk temperature, $kT_{\text{s}}$, the corona temperature, $kT_{\text{e}}$, and the corona size, $L$; the latter is always large, between 5600~km and 7200~km.

\begin{table}\renewcommand\arraystretch{1.2}
    \caption{Spectral parameters of the joint fitting for \maxi. The error indicates the 1-$\sigma$ confidence level. See the text for more details about the parameters.}
    \centering
    \begin{threeparttable}
    \begin{tabularx}{\columnwidth}{cccc} 
        \hline
        Component & Parameter & Model 1 & Model 2 \\
        \hline
        \texttt{TBfeo} & $N_{\text{H}}$ ($10^{22}$~cm$^{-2}$) & $4.18\pm 0.08$ & $4.18\pm 0.08$ \\
                       & $A_{\text{O}}$ & $0.44\pm 0.05$ & $0.52\pm 0.05$ \\
                       & $A_{\text{Fe}}$ & $1.74\pm 0.08$ & $1.51\pm 0.09$ \\
        \texttt{diskbb} & $kT_{\text{in}}$ (keV) & $1.138\pm 0.002$ & $1.139\pm 0.002$ \\
        \texttt{vKompthdk} & $kT_{\text{s}}$ (keV) & $0.59\pm 0.06$ & $0.59\pm 0.07$ \\
                           & $kT_{\text{e}}$ (keV) & $330\pm 50$ & $330\pm 30$ \\
                           & $\Gamma$ & $2.91\pm 0.06$ & $2.88\pm 0.04$ \\
                           & $L$ ($10^{3}$~km) & $6.5\pm 0.4$ & $6.5\pm 0.5$ \\
                           & $\eta^{\text{a}}$ & $1^{}_{-0.07}$ & $1^{}_{-0.07}$ \\
                           & $\delta\dot{H}$ & $0.15\pm 0.02$ & $0.16\pm 0.02$ \\
        \texttt{gaussian1} & LineE (keV) & $6.80\pm 0.04$ & $6.75\pm 0.05$ \\
                           & $\sigma$ (keV) & $0.96\pm 0.04$ & $1.01\pm 0.05$ \\
                           & Norm ($10^{-2}$) & $6.2\pm 0.5$ & $5.9\pm 0.4$ \\
        \texttt{gaussian2} & LineE (keV) & $^{}_{}$ & $6.62\pm 0.02$ \\
                           & $\sigma$ (keV) & $^{}_{}$ & $0.14\pm 0.02$ \\
                           & Norm ($10^{-3}$) & $^{}_{}$ & $3.0\pm 0.7$ \\
        \hline
        \multicolumn{2}{c}{$\chi^{2}/\text{bin}$ (SSS$^{\text{b}}$)} & 4137.84 / 4576 & 4098.42 / 4576 \\
        \multicolumn{2}{c}{$\chi^{2}/\text{bin}$ (rms$^{\text{c}}$)} & 46.76 / 5 & 45.81 / 5 \\
        \multicolumn{2}{c}{$\chi^{2}/\text{bin}$ (phase lag$^{\text{d}}$)} & 15.57 / 5 & 15.47 / 5 \\
        \multicolumn{2}{c}{$\chi^{2}/\text{d.o.f.}$} & 4200.08 / 4520 & 4159.70 / 4517 \\
        \hline
        \multicolumn{2}{c}{Disk flux$^{\text{e}}$ (erg cm$^{-2}$ s$^{-1}$)} & \multicolumn{2}{c}{$1.577\pm 0.004 \times 10^{-7}$} \\
        \multicolumn{2}{c}{Corona flux$^{\text{e}}$ (erg cm$^{-2}$ s$^{-1}$)} & \multicolumn{2}{c}{$0.444\pm 0.003 \times 10^{-7}$} \\
        \hline
    \end{tabularx}\label{tab:spectral_par}
    \begin{tablenotes}
        \item[a] We give only the negative error of the feedback fraction, $\eta$, because the parameter is consistent with 1, the maximum possible value at the 1-$\sigma$ confidence level.
        \item[b] Steady-state spectrum.
        \item[c] Fractional rms spectrum.
        \item[d] Phase-lag spectrum.
        \item[e] Unabsorbed flux.
    \end{tablenotes}
    \end{threeparttable}
\end{table}

\section{Discussion}\label{sec:discussion}

We have analyzed 36 \nicer\ observations of the black hole candidate \maxi\ during the 2017/2018 outburst. We identify a type-B QPO in 26 \nicer\ orbits where the source is in the SIMS, during which the inner disk is relatively hot, $kT_{\text{in}}\sim 1.1$~keV, and the corona photon index is relatively high, $\Gamma\sim 2.9$. From the simultaneous fit of the energy spectra of the source and the rms and phase-lag spectra of the type-B QPO, we find that the corona has a size $L=6500\pm 500$~km and a feedback fraction $\eta=1_{-0.07}$\footnote{This value of the feedback fraction gives $\eta_{\text{int}}=0.33\pm 0.02$ which means that $\sim 33\%$ of the total Comptonized photons impinge back onto the disk, while the remaining 67\% are the observed Comptonized photons.}. We detect an additional significant narrow iron line with central energy of $\sim 6.6$~keV and FWHM of $\sim 0.1$~keV, consistent with a large corona illuminating the outer and cooler parts of the accretion disk.

\subsection{The phase-lag spectrum of type-B QPOs}

By systematically studying the type-B QPOs using \rxte\ data, the lags of type-B QPOs are hard for the BHB systems of low inclination angle, while they are either hard or soft for the BHB systems of high inclination angle~\citep{2017MNRAS.464.2643V,2017MNRAS.466..564G}. The hard lags are likely due to inverse-Compton scattering of the soft disk photons in the corona~\citep{2008A&A...489..481K}, while the soft lags are likely due to the down-scattering of the hard photons in the disk~\citep{2014A&ARv..22...72U}. \citet{2016MNRAS.460.2796S} studied the hard lags of type-B QPOs in GX~339$-$4 using \rxte\ data and suggested that a corona with a large-scale height ($\sim 1.8\times 10^4$~km) is the jet base. This scenario of a large corona derived from the phase-resolved spectroscopy of the type-B QPO is later modeled by~\citet{2020A&A...640L..16K} who proposed that the type-B QPO in GX~339$-$4 originates from a precessing jet. Note that \rxte\ is not sensitive to photons below 2--3~keV. Using \nicer\ observations of MAXI~J1348$-$630, \citet{2020MNRAS.496.4366B} found that the phase lags are positive both in the 3--10~keV and the 0.7--2~keV with respect to the reference band at 2--3~keV. The positive phase lags in the 0.7--2~keV band exclude the possibility that the type-B QPOs are due to propagation of mass accretion rate fluctuations which, in this scenario, the phase lags below 2~keV relative to the 2--3~keV band should be negative. Comptonization of the soft photons in the corona naturally explains the positive lags at energy above 3~keV, while~\citet{2020MNRAS.496.4366B} suggested that the positive lags of the type-B QPO below 2~keV in MAXI~J1348$-$630 could be due to Compton down-scattering in the corona by using Monte Carlo simulations assuming a flat seed spectrum between 2 and 3~keV. However, because their seed spectrum does not emit below 2~keV, they neglected the effect of the direct emission of the seed source on the phase-lag spectrum. Indeed, the direct emission of a more realistic seed source (e.g.\ an accretion disc) leads to a flat phase-lag spectrum below $\sim$ 3~keV~\citep[Kylafis, priv. comm.; see also Figure 1 in][]{2022MNRAS.515.2099B}, contrary to the observations.

From the phase-lag spectrum (right panel of Fig.~\ref{fig:rms_and_plag_spect}) in \maxi, the phase lags generally decrease as the energy increases, with the lags being a minimum at around 6~keV. If we take the lowest energy band (1--2.5~keV) as the reference band, in \maxi\ all the lags are soft. \citet{2018ApJ...865L..15S} proposed that the soft lags of the type-B QPOs in \maxi\ are due to the phase offset between the peaks in the corona emission and the modulation of the disk spectrum. Based on the fitting results of our \texttt{vKompthdk} model, we find that $33\%\pm 2\%$ of the corona photons return to and are reprocessed in the accretion disk, producing the soft lags. Therefore, the observed soft lags can be explained as the light-crossing time of a large corona illuminating the disk. Note that the minimum in the phase-lag spectrum at $\sim$ 6~keV could be related to the iron line feature, but since the type-B QPO in this source is weak, we do not have good enough resolution to perform a detailed line analysis.

\subsection{Comparison of corona models of variability}

X-ray variability in the accreting X-ray binaries is generally classified as broadband noise and QPOs~\citep[][for a review]{2019NewAR..8501524I}. The broadband noise at low frequencies usually shows large hard lags that are thought to be due to Comptonization, while at relatively high frequencies it sometimes shows soft lags, which have been proposed to be caused by X-ray reverberation of corona photons reflected off the accretion disk~\citep{2014A&ARv..22...72U}. In a lamppost geometry of the corona above a black hole, the model \texttt{reltrans} calculates the difference in the light-travel time of the photons reflected off the accretion disk relative to the corona photons that travel directly to the observer~\citep{2019MNRAS.488..324I}. This model is able to fit the time-lag spectrum of the broadband noise in the black hole X-ray transient MAXI~J1820+070, giving a corona height of up to $\sim$ 500~$R_{\text{g}}$~\citep{2021ApJ...910L...3W}, equivalent to $\sim$ 6400~km for an 8.5-$M_{\odot}$ black hole~\citep{2020ApJ...893L..37T}. The lamppost geometry of the corona is a simplification to allow the calculation of the ray tracing in the spacetime around the black hole. On the other hand, this model cannot explain the hard lags observed sometimes in these systems that are therefore assigned to either Comptonization or fluctuation of accretion propagation~\citep{2006MNRAS.367..801A,2008A&A...489..481K}. Since the soft (reverberation) and the hard (inverse-Compton scattering and mass accretion rate fluctuation) are treated separately, the model is in essence two separate mechanisms that are connected by the fitting procedure.

From the perspective of QPOs that dynamically originate from the L-T precession, \citet{2016MNRAS.461.1967I} and \citet{2022MNRAS.511..255N} developed a tomographic model that fits the QPO phase-dependent energy spectrum, explaining the energy shifts of the observed iron $K_{\alpha}$ emission line in different QPO phases. Using \nicer\ and \textit{NuSTAR} data of GRS~1915+105 with a QPO frequency at 2.2~Hz, \citet{2022MNRAS.511..255N} measured a thermalization time delay of 70~ms, which is too long since in this long time delay the QPO signal would be washed out. \citet{2022MNRAS.511..255N} attributed this problem to the oversimplification of the precessing corona model. For instance, the authors did not take into account the precessing corona/jet obscuring different disk azimuths, which would result in a variation of the shape of the observed disk spectrum, and they ignored the light-crossing delays which can be of the order of milliseconds. Systematic error in the spectral modeling would potentially affect the measured time lags as well. \citet{2022MNRAS.511..255N} also measured an inner radius, $R_{\text{in}}=1.5 R_{\text{g}}$, of the accretion disk which is too small to produce the 2.2-Hz QPO predicted by the L-T precession of the corona lying inside a truncated disk~\citep{2009MNRAS.397L.101I}. It could be that the corona is not horizontally, but more likely vertically, extended in the form of a jet-like structure, namely the outflow~\citep{2016MNRAS.460.2796S,2020A&A...640L..16K}.

Regardless of the dynamical origin of the QPOs, the time-dependent Comptonization model \texttt{vKompthdk} developed by~\citet{2020MNRAS.492.1399K} and~\citet{2022MNRAS.515.2099B} describes both the hard and soft lags by considering inverse-Compton scattering in the corona and thermal reprocessing in the disk. Since the model solves the linearized Kompaneets equation, it provides the distribution of photons in energy at any given time, regardless of the history of these photons via a single mechanism\footnote{In that respect this model is simpler than the reverberation model that requires two separate mechanisms to explain both the soft and hard lags at different Fourier frequencies.}. This model can successfully fit the steady-state Comptonization spectrum and the rms and phase-lag spectra of QPOs with energies in the 1--100~keV range~\citep{2020MNRAS.492.1399K,2021MNRAS.501.3173G,2022NatAs...6..577M,2022MNRAS.513.4196G,2022MNRAS.512.2686Z}. A limitation of the model is that even though it considers the thermal reprocessing of hard photons in the accretion disk, it ignores the relativistic reflection that produces the atom florescent emission lines and Compton hump~\citep{2014ApJ...782...76G}. The corona size measured by \texttt{vKompthdk} is generally big compared with the corona size predicted by L-T precession inside the truncated disk (see supplementary Fig.~4 in~\citealt{2022NatAs...6..577M} and Fig.~5 in~\citealt{2022MNRAS.513.4196G}). If the corona size is large, regardless of the feedback fraction, a big corona is likely to be vertically extended~\citep{2022NatAs...6..577M,2022MNRAS.512.2686Z}. A dual-Comptonization model is also used to fit the \nicer\ data of the type-B QPO of MAXI~J1348$-$630~\citep{2021MNRAS.501.3173G}, showing that the inner part of the corona is small ($L\simeq 400$~km) and spherical or slab, with the hard photons efficiently returning to the accretion disk, while the outer part of the corona is large ($L\simeq 10^{4}$~km) and jet-like with nearly no hard photon feedbacking onto the accretion disk. The corona geometry, e.g.\ the size, measured from different models mentioned above has features in common (see below).

\begin{figure*}
    \includegraphics[width=0.59\textwidth]{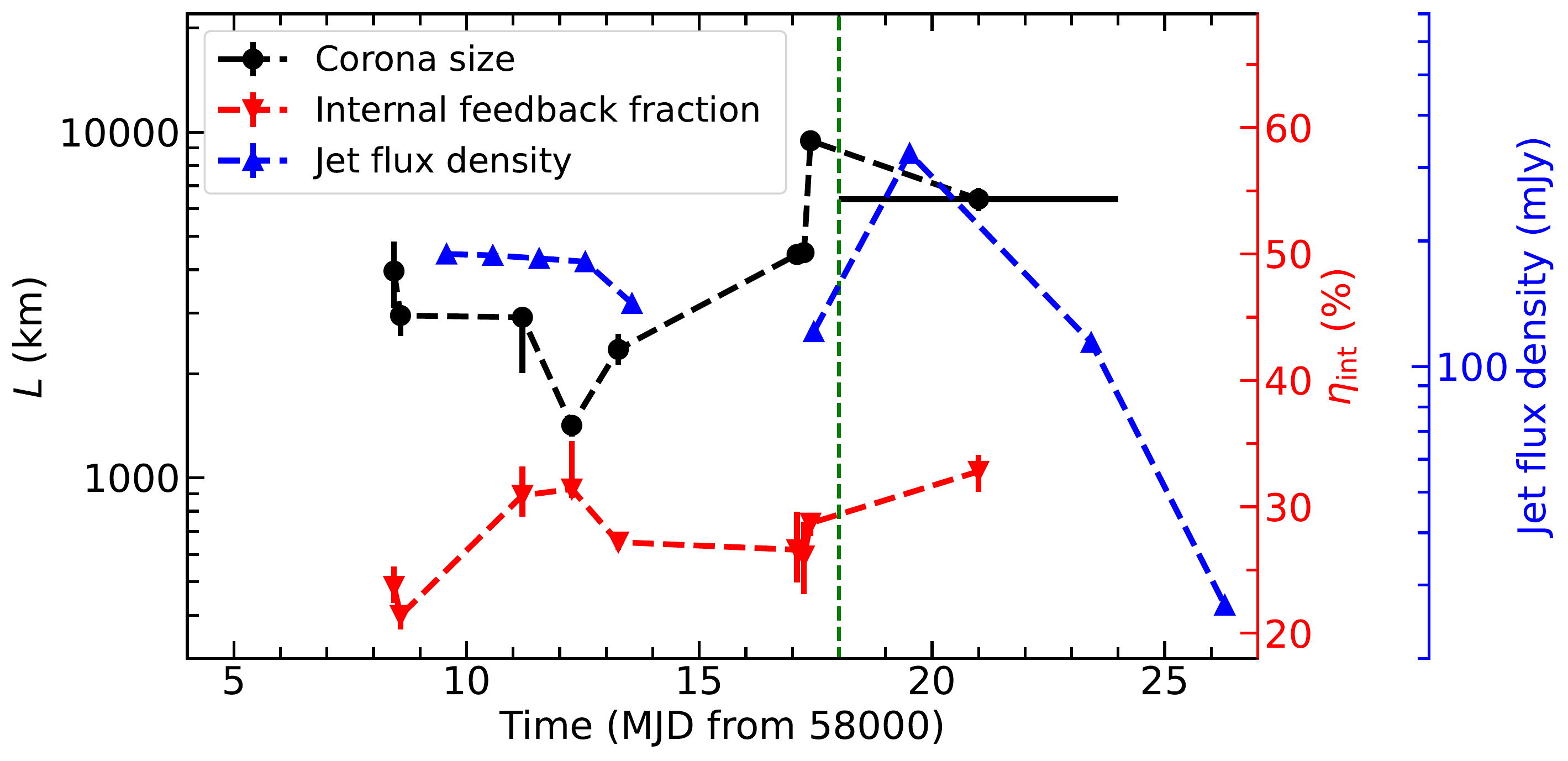}
    \includegraphics[width=0.39\textwidth,trim=-20 -100 20 0]{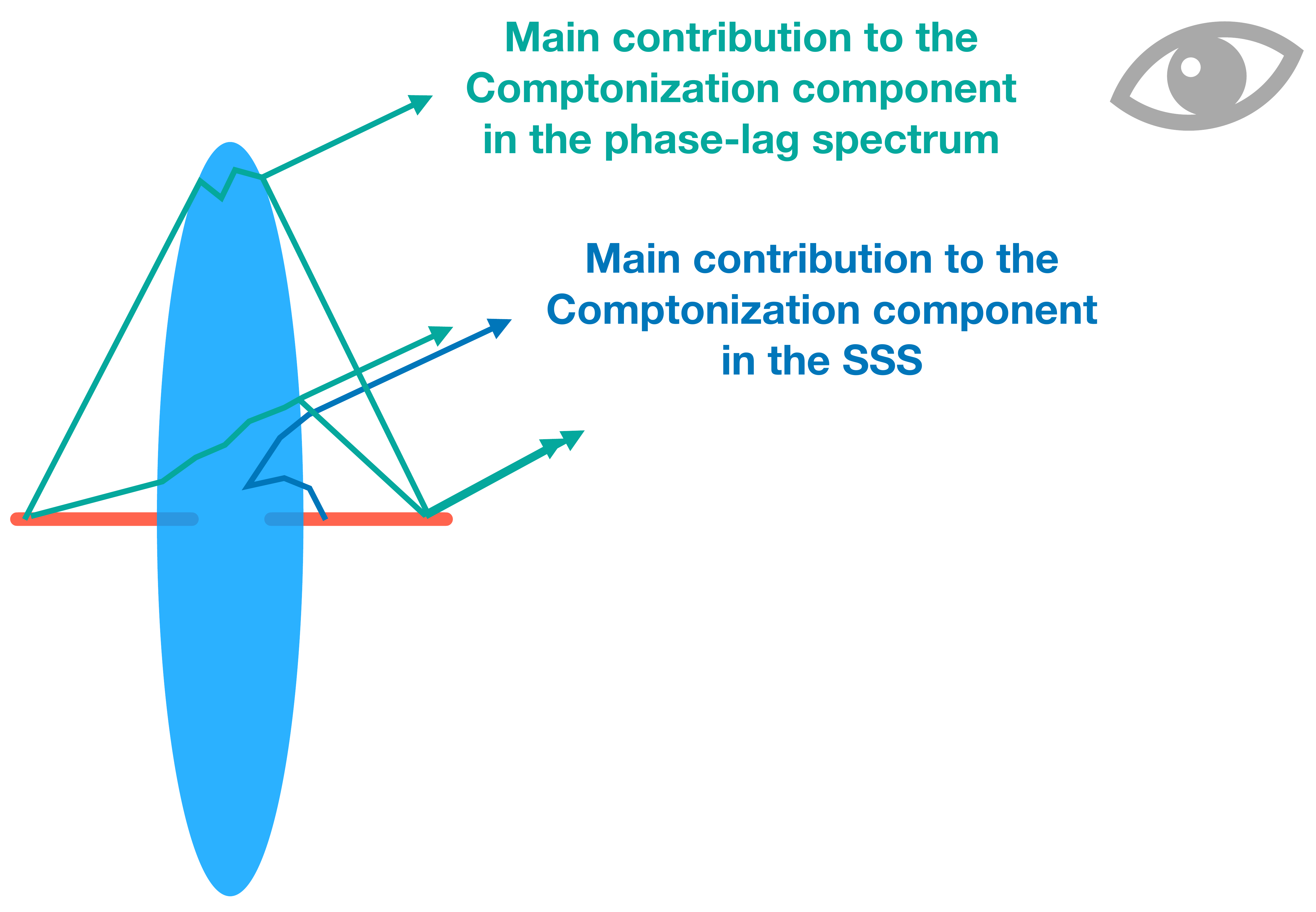}
    \caption{Left panel: The evolution of the corona geometry and the jet flux from the HIMS to the SIMS in \maxi. The black, red, and blue points indicate the corona size, the intrinsic feedback fraction, and the jet flux density, respectively. The blue dashed line is discontinuous since the jet is quenched from MJD~58009 to 58017. The green dashed line indicates the time of the transition from the HIMS to the SIMS. Right panel: A schematic figure of the corona geometry of \maxi\ in the SIMS. The inner part of the disk provides the seed photons to the Comptonized emission that dominates the steady-state energy spectrum of the source, while the outer parts of the disk provide the seed photons to the Comptonized emission that dominates the rms and phase-lag spectra of the QPO. See text for more details.} 
    \label{fig:corona}
\end{figure*}

\subsection{Corona geometry}

Fast transitions between type-B and other types (A and C) QPOs are often observed~\citep[e.g.][]{2004A&A...426..587C,2021MNRAS.505.3823Z}. The detailed spectral-timing analysis of H~1743$-$322 indicates that the transition from type-B to type-C QPOs could be explained by the presence of a jet or a vertically extended optically thick Comptonization region~\citep{2022MNRAS.tmp.2333H}. From a spectral-timing analysis of the \hxmt\ data of MAXI~J1348$-$630 in the SIMS, \citet{2022ApJ...938..108L} proposed a vertically extended corona that is the base of the jet, and explained the disappearance and appearance of the type-B QPO as the jet being parallel to the BH spin axis or not, due to the Bardeen-Petterson effect. In our study of \maxi\ in the SIMS, we notice that the type-B QPOs disappear in some \nicer\ orbits as shown in Fig.~\ref{fig:hid} which may be explained by the L-T precession of the corona or the jet proposed by~\citet{2022ApJ...938..108L}.

The modeling of the corona through type-B QPOs in MAXI~J1348$-$630 shows that the size of the jet-like corona is $\sim 10^{4}$~km~\citep{2021MNRAS.501.3173G}. This size is comparable with the corona size of $6500\pm 500$~km that we find in \maxi. The intrinsic feedback fraction in this work is $33\%\pm 2\%$, which indicates that the corona should be covering the accretion disk to some extent, as shown in the right panel of Fig.~\ref{fig:corona}. From the fitting results in Tab.~\ref{tab:spectral_par}, the inner disk provides the seed photons and the Comptonized photons contribute mainly to the steady-state energy spectrum, while the outer disk provides relatively cold seed photons and the long light-crossing time of the Comptonized photons contribute mainly to the phase-lag spectrum. In \maxi\ we have already measured the corona geometry through type-C QPOs in the HIMS from MJD~58008 to 58017~\citep{2022MNRAS.512.2686Z}. The left panel of Fig.~\ref{fig:corona} shows the corona size, intrinsic feedback fraction, and the 9-GHz jet flux density~\citep{2019ApJ...883..198R} using the results both in~\citet{2022MNRAS.512.2686Z} and this work\footnote{In this work, we improve the fitting model compared to~\citet{2022MNRAS.512.2686Z}. For more information about the refined model, see subsection~\ref{subsec:joint}.}. Note that we have converted the feedback fraction, $\eta$, into the intrinsic feedback fraction, $\eta_{\text{int}}$. As discussed in ~\citet{2022MNRAS.512.2686Z}, the corona on MJD~58017 is a vertically extended, jet-like corona, which expands vertically to its maximal size, $L\sim 9300$~km, with $\eta_{\text{int}}\sim 28\%$, two days before the transient jet reaches the maximum flux density. After MJD~58017, \maxi\ transits into the SIMS. In the SIMS from MJD~58018 to 58024, we measure a corona size $L\sim 6500$~km with $\eta_{\text{int}}\sim 33\%$. Compared to the corona geometry in the end of the HIMS on MJD~58017, in the SIMS the corona size contracts and the hard photons feedback onto the disk more efficiently, indicating that the corona contracts vertically and expands horizontally. After MJD~58024 no QPO appears in the intermediate and high-soft states and the transient jet flux density gradually decays until it is no longer detected. The change of the morphology of the corona geometry in the SIMS, together with that in the HIMS, and the change of the jet flux density, suggests that the increasing size of the jet-like corona may give rise to the large-scale transient jet ejecta lagging behind the change of the corona size. After the ejection, the corona in the jet base contracts and the transient jet loses its energy source so the observed radio flux density drops. This suggests a connection between the corona and the jet.

The corona-jet connection has been investigated using \texttt{vKompthdk} through type-C QPOs in GRS~1915+105 with 12-year \rxte\ observations~\citep{2021MNRAS.503.5522K,2022NatAs...6..577M,2022MNRAS.513.4196G}. The connection between the corona and the jet in the persistent source GRS~1915+105 is similar to that in the BH X-ray transient \maxi. In GRS~1915+105, when the radio emission is weak the corona covers large parts of the accretion disk and the hard photons efficiently feedback onto the disk, while when the radio emission is strong the corona is vertically extended and jet-like, and less hard photons feedback onto the disk~\citep{2022NatAs...6..577M}. The only difference is that, in the SIMS of \maxi, the hard photons feedback onto the disk more efficiently than in the HIMS. The $\sim 33\%$ intrinsic feedback fraction in the SIMS of \maxi\ is similar to the $\sim 35\%$ intrinsic feedback fraction in the SIMS of MAXI~J1348$-$630~\citep{2021MNRAS.501.3173G,2022MNRAS.515.2099B}. In fact the feedback fraction of hard photons may either decrease or increase, depending on the balance between the Compton cooling process in the disk and the heating up of the corona~\citep{2001MNRAS.321..549M,2020MNRAS.492.1399K}.

The evolution of the corona height using the light-crossing delays in reverberation shows quite a similar trend to our measurements using \texttt{vKompthdk} through the QPOs. \citet{2022ApJ...930...18W} studied archival data of 10 black hole candidates with \nicer\ and found that during the hard to intermediate state transition the soft lags first decrease and then increase~\citep[see Fig.~6 in][]{2022ApJ...930...18W}. Assuming a lamppost geometry, the corona height in the hard state decreases from $\sim 1000$~km to $\sim 200$~km, in the HIMS it increases monotonically up to 9000~km, and in the SIMS it decreases slightly to $\sim 6000$~km. Comparing these results with those in the left panel of Fig.~\ref{fig:corona}, we notice a general consistence of the measurements of the corona size. This consistence suggests the possibility to combine \texttt{reltrans} and \texttt{vKompthdk} to measure the corona geometry when reverberation lags and QPOs appear simultaneously.



\section*{Acknowledgements}

We thank the referee for constructive comments that helped us improve the quality of this paper. YZ acknowledges support from the China Scholarship Council (CSC~201906100030). MM acknowledges support from the research programme Athena with project number 184.034.002, which is (partly) financed by the Dutch Research Council (NWO). FG is a CONICET researcher. FG acknowledges support by PIP~0113 (CONICET) and PIBAA~1275 (CONICET). This work received financial support from PICT-2017-2865 (ANPCyT). DA acknowledges support from the Royal Society. TMB acknowledges financial contribution from PRIN INAF~2019 n.15. RM acknowledges support from the China Scholarship Council (CSC~202104910402). MM, FG and TMB have benefited from discussions during Team Meetings of the International Space Science Institute (Bern), whose support we acknowledge.


\section*{Data Availability}

The X-ray data used in this article are accessible at NASA's High Energy Astrophysics Science Archive Research Center~\url{https://heasarc.gsfc.nasa.gov/}. The software GHATS for Fourier timing analysis is available at~\url{http://www.brera.inaf.it/utenti/belloni/GHATS_Package/Home.html}. The time-dependent Comptonization model and the generator of the MCMC corner plot are available at the GitHub repositories~\url{https://github.com/candebellavita/vkompth} and~\url{https://github.com/garciafederico/pyXspecCorner}.



\bibliographystyle{mnras}
\bibliography{reference} 



\appendix

\section{Observations and orbits used for concatenation}

\begin{table}\renewcommand\arraystretch{1.2}
    \caption{\nicer\ observations and QPO frequency of \maxi. The error bars indicate the 1-$\sigma$ confidence level.}
    \centering
    \begin{tabularx}{\columnwidth}{ccccc} 
        \hline
        Observation ID & Orbit & MJD start & QPO frequency (Hz) \\ 
        \hline
        1050360114 & 1 & 58018.88 & $6.1\pm 0.1$ \\ 
                   & 2 & 58018.94 & $6.7\pm 0.4$ \\ 
        1050360115 & 1 & 58019.01 & $6.2\pm 0.4$ \\ 
                   & 4 & 58019.22 & $5.6\pm 0.4$ \\ 
                   & 5 & 58019.28 & $5.5\pm 0.5$ \\ 
                   & 6 & 58019.47 & $5.7\pm 0.3$ \\ 
                   & 7 & 58019.54 & $5.5\pm 0.3$ \\ 
                   & 8 & 58019.60 & $5.3^{+0.6}_{-0.1}$ \\ 
                   & 13& 58019.92 & $5.2\pm 0.5$ \\ 
                   & 14& 58019.99 & $6.0\pm 0.2$ \\ 
        1050360116 & 1 & 58020.05 & $5.6\pm 0.5$ \\ 
                   & 4 & 58020.25 & $5.1\pm 0.2$ \\ 
                   & 5 & 58020.31 & $5.5\pm 0.4$ \\ 
                   & 6 & 58020.37 & $5.8\pm 0.2$ \\ 
                   & 8 & 58020.50 & $6.2\pm 0.2$ \\ 
        1050360117 & 2 & 58021.15 & $6.2\pm 0.4$ \\ 
                   & 3 & 58021.40 & $5.3\pm 0.2$ \\ 
                   & 4 & 58021.60 & $5.8\pm 0.3$ \\ 
                   & 5 & 58021.77 & $5.5^{+0.1}_{-0.5}$ \\ 
        1050360118 & 1 & 58022.24 & $6.0\pm 0.2$ \\ 
        1050360119 & 1 & 58023.26 & $6.1\pm 0.1$ \\ 
                   & 3 & 58023.77 & $5.1\pm 0.1$ \\ 
        1050360120 & 1 & 58020.57 & $6.6\pm 0.4$ \\ 
                   & 2 & 58020.63 & $5.8\pm 0.2$ \\ 
                   & 4 & 58020.76 & $5.6\pm 0.2$ \\ 
                   & 7 & 58020.95 & $4.8\pm 0.1$ \\ 
        \hline
    \end{tabularx}\label{tab:obs}
\end{table}



\bsp	
\label{lastpage}
\end{document}